\documentclass[preprint,12pt]{elsarticle}

\makeatletter
\newcommand{\labitem}[2]{%
\def\@itemlabel{\textbf{#1}}
\item
\def\@currentlabel{#1}\label{#2}}
\makeatother
\usepackage{amssymb}

\usepackage{booktabs}

\usepackage[T1]{fontenc}
\usepackage[utf8]{inputenc}
\usepackage{url}
\usepackage{tikz}
\graphicspath{{figures/}}

\def\chm{\tikz\fill[scale=0.4](0,.35) -- (.25,0) -- (1,.7) -- (.25,.15) -- cycle;} 

\journal{Arxiv.org}

\begin{document}

\begin{frontmatter}



\title{Kooplex: collaborative data analytics portal for advancing sciences}

\author[ad1,ad3]{D. Visontai}
\author[ad1,ad3]{J. Stéger}
\author[ad2,ad3]{J. M. Szalai$-$Gindl}
\author[ad1,ad3]{L. Dobos}
\author[ad1,ad3]{L. Oroszl{\'a}ny}
\author[ad1,ad3]{I. Csabai}
\address[ad1]{Department of Physics of Complex Systems, E{\"o}tv{\"o}s Lor{\'a}nd University, H-1117, Pázmány Péter sétány 1/a. Budapest, Hungary}
\address[ad2]{Department of Information Systems, E{\"o}tv{\"o}s Lor{\'a}nd University, H-1117, Pázmány Péter sétány 1/c. Budapest, Hungary}
\address[ad3]{Department of Computational Sciences, Wigner Research Centre for Physics of the HAS,
Konkoly-Thege Miklós út 29-33., Budapest 1121, Hungary}

\newcommand{\kooplex}{\texttt{KOOPLEX}}

\begin{abstract}
The web of collaborations between individuals and group of researchers continuously grows thanks to online platforms, where people can share their codes, calculations, data and results. These virtual research platforms are innovative, mostly browser-based, community-oriented and flexible.  They provide a secure working environment required by modern scientific approaches. There is a wide range of open source and commercial solutions in this field and each of them emphasizes the relevant aspects of such a platform differently.

In this paper we present our open source and modular platform, \kooplex\footnote{https://kooplex.github.io}, that
combines such key concepts as dynamic collaboration, customizable research environment, data sharing, access to datahubs, reproducible research and reporting. 
It is easily deployable and scalable to serve more users or access large computational resources.
\end{abstract}


\end{frontmatter}


\newcommand{\user}{\emph{User}}
\newcommand{\users}{\emph{Users}}
\newcommand{\notebook}{\emph{Notebook}}
\newcommand{\notebooks}{\emph{Notebooks}}
\newcommand{\report}{\emph{Report}}
\newcommand{\reports}{\emph{Reports}}
\newcommand{\volume}{\emph{Volume}}
\newcommand{\volumes}{\emph{Volumes}}
\newcommand{\functionalvolume}{\emph{Functional Volume}}
\newcommand{\functionalvolumes}{\emph{Functional Volumes}}
\newcommand{\storagevolume}{\emph{Storage Volume}}
\newcommand{\storagevolumes}{\emph{Storage Volumes}}
\newcommand{\group}{\emph{Group}}
\newcommand{\groups}{\emph{Groups}}
\newcommand{\collaborator}{\emph{Collaborator}}
\newcommand{\collaborators}{\emph{Collaborators}}
\newcommand{\public}{\emph{Public}}
\newcommand{\project}{\emph{Project}}
\newcommand{\projects}{\emph{Projects}}
\newcommand{\scope}{\emph{Scope}}
\newcommand{\scopes}{\emph{Scopes}}
\newcommand{\container}{\emph{Container}}
\newcommand{\containers}{\emph{Containers}}
\newcommand{\service}{\emph{Service}}
\newcommand{\services}{\emph{Services}}
\newcommand{\home}{\emph{Home}}
\newcommand{\share}{\emph{Share}}
\newcommand{\workdir}{\emph{Workdir}}
\newcommand{\gitfolder}{\emph{Vc}}
\newcommand{\seafile}{\emph{Cloud}}


\newcommand{\kooplex}{\texttt{KOOPLEX}}
\newcommand{\kooplexhub}{\texttt{KOOPLEX Hub}}
\newcommand{\kooplexconfig}{\texttt{KOOPLEX Config}}
\section{Introduction}
\label{intro}
Innumerable fields of research, government funded or commercial services provide TBs or PBs of data annually. These public or partially public datasets residing in datahubs give significant momentum to several disciplines and to researchers.
Virtual research environments \cite{candela2013virtual} (VRE) are web-based, flexible, community-oriented and secure working environments, through which users can access these data and additional computational resources needed for the production of meaningful results.
VREs relieve users from much of the technical burdens such as software installations, complicated file and user management and general maintenance. 
For such a VRE to be attractive the platform's layout and user interface should satisfy the needs of the users and suit their skills. Users can have very different level of technical skillsets and many of the commercial platforms target consumers with ready-made and standardized workflows and visualization tools.

There are platforms that were developed specifically either for a datahub \cite{SOILLE201830} and others that were customized for the inner structure of a given institute.

Commercial platforms have the disadvantage from a researcher's point of view that the extent of collaboration is limited: others can't join without paying the licence fee for its usage. Also every customization, that involves development on the platform is cost sensitive.

Data sharing and reproducibility is another aspect. There are many ways and solutions for data sharing, but sharing of the data analysis process in a reproducible way is somewhat lagging behind. 
Scientific results are traditionally published as human readable articles, but human language lacks the precision and details of computer codes, therefore reproducing results based on the written information, even in possession of the data, is often a long and tiresome process, if possible at all. 

The figures in traditional articles without the ability for zooming or subsetting also hide many of the details present in the data and especially multidimensional data sets are hard to represent in passive two dimensional prints. In the late 80s Mathematica’s \footnote{Wolfram Research Inc., https://www.wolfram.com/mathematica/} first notebook frontend was released. Since then slowly but steadily other languages like  R  and Python picked up the concept of “reproducible research” \cite{munafo2017manifesto}. In the last couple of years R-Studio\footnote{Rstudio, https://www.rstudio.com} and Jupyter Notebooks\footnote{JupyterHub, https://github.com/jupyterhub/jupyterhub} \cite{kluyver2016jupyter} - capable to handle Python \cite{4160251}  and several other languages - became a standard for data analytics and visualization. Notebooks integrate the analysis and visualization process and produce output that can be rendered as rich interactive web pages. 


The paper is organized as follows: In \textbf{Section \ref{kooplexintro}.} we introduce our prototype platform or VRE, the \kooplex and in \textbf{Table \ref{table:terms}.} we define how terms are used in \kooplex.
In \textbf{Section \ref{sec:hub}. and \ref{sec:config}.} we describe the architecture of the \kooplex{} platform.
In \textbf{Section \ref{sec:platform-comparison}}. we define aspects that are relevant when choosing a multiuser platform for datascience or business intelligence. It also  allows us to give context for \kooplex. 
And finally we present some use cases of this platform in \textbf{Section \ref{sec:kooplex-usecase}}.


\section{\kooplex{} architecture}
\label{kooplexintro}
With \kooplex{} we intended to add many features that enhances dynamic collaboration between users with any kind of institutional background and every user's wish to work in their customized/habitual environment.
It should also be emphasized, that in many cases the ability of testing out the latest version of a software and developing someone's own code in a secluded environment is crucial for efficient research.
With nowadays virtualization techniques it is possible to integrate well established tools along with experimental ones into an easily deployable platform. 
\kooplex{} is therefore a modular platform that accommodates several services. \users{} can work in \notebooks, it provides access to computational resources and supports the creation of reports in various formats. \users{} may manage and collaborate within projects, develop and create workflows, and publish results.
 
The \kooplex{} platform consists of two parts, the \kooplexconfig\footnote{\url{https://github.com/kooplex/kooplex-config}}  and the \kooplexhub\footnote{\url{https://github.com/kooplex/kooplex-hub}}.

\kooplexconfig{} defines the structure of the virtualized backend services and orchestrate the installation of the platform. Any new services/modules are added via the \kooplexconfig. 

The \kooplexhub{} is a django based frontend/webserver of the platform. Users may access \containers{} and services on the 

There is a demo site, which can be accessed following the \kooplex{} website \url{https://kooplex.github.io/}.

In \textbf{Table \ref{table:terms}.} we define how terms are used in \kooplex.

\begin{table}[h!]
\label{table:terms}
\caption{Usage of terms in this paper and \kooplex}
    \begin{tabular}{p{2.0cm} p{10cm}}
        \addlinespace[0.1cm]
        \hline
        Terms &  Description  \\
        \hline  \hline
        \addlinespace[0.2cm]

        \user & stands for a registered actor.
Typically it is a person but it can be a script created by a person as well. \\[.5em]
        

        \collaborator & users are collaborators if they are associated with the same project, sharing the same workspace or data. \\[.5em]


        \project & is a unique basket of resources granted to \users.
It offers collaboration, different ways of data sharing and a default common environment. \\[.5em]
        
        \scope & defines the relation between a \project{} or a \report{} and a \user. A \project{} or a \report{} can be \emph{private}, \emph{internal} or \emph{public}. A \user{} can be the owner, an administrator or a collaborator of a project. \\[.5em]
        
        \notebook & this could refer to any browser based application such as Jupyter notebook server, R Studio server, bash terminal emulator etc.  \\[.5em]
        
        \container & always refers to Docker container, which runs a \notebook.  It is a virtual machine in which one can access data, write and run code and visualize results. \\[.5em]
        
        \volume &  always refers to Docker volumes. A docker volume is an interface for storages, let it be a directory or a disk on a remote server. It facilitates their usage and hides the technical details.
We distinguish two types: a \storagevolume{} is where large amount of data is stored, and \functionalvolume{} is for installed packages and softwares, scripts or sample data, that is intended for other users as well. These \volumes{} can be dynamically mounted into a \container{} as needed.  \\[.5em]
        
        \report & is a scientific output of a \user{}. It consists of an image, a set of codes (e.g. Jupyter notebook), datafiles etc. converted into either a static page with or without interaction or a hosted application. \\[.5em]
        
        \service & an indepently developed code or software, which we interface to \kooplex{}. \\
    \end{tabular}
\end{table}

\section{The \kooplexhub}
\label{sec:hub}

\begin{figure}[h!]
    \centering
    \includegraphics[width=400px]{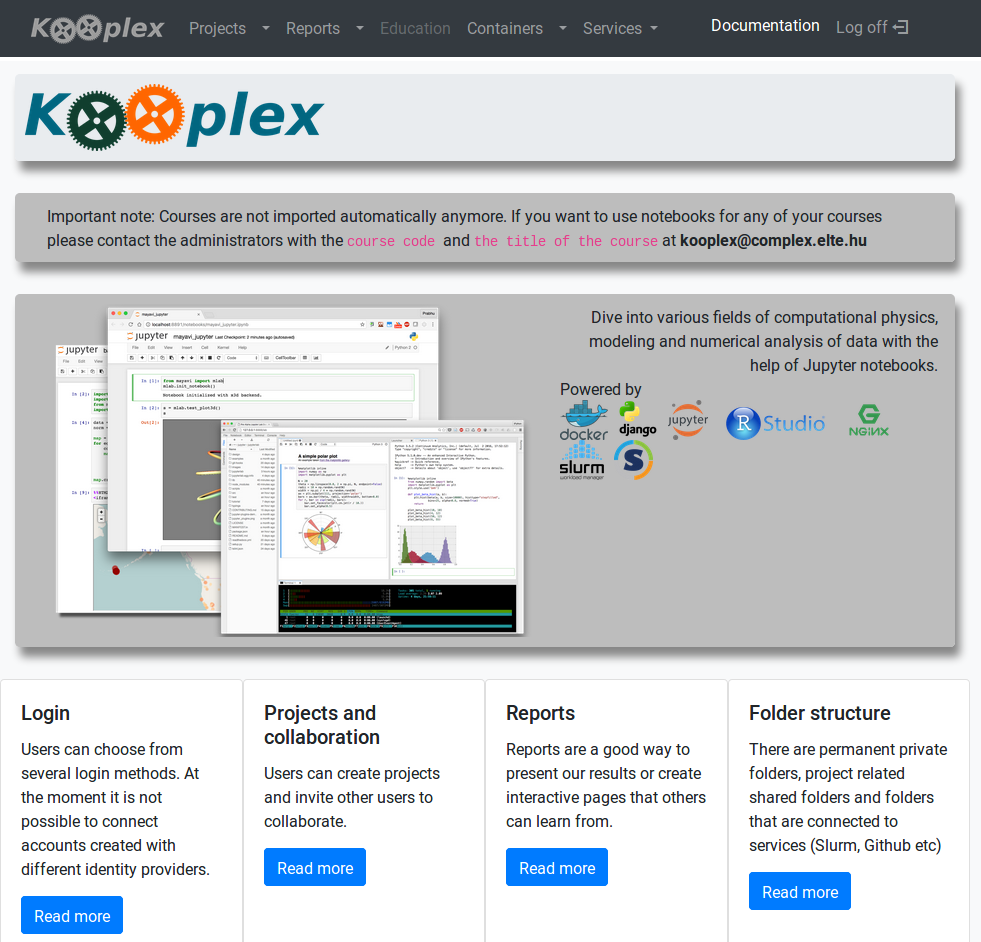}
    \caption{The frontpage \kooplexhub{} allows users to create projects, set level of collaboration, create containers and allocate software modules and storages to it. Reports can be finalized. Under the education tag courses can be defined to which students can join.}
    \label{fig:kooplex_frontend}
\end{figure}

The \kooplexhub{} is a Django based webserver that connects all the components and maintains the consistency of databases, the filesystem and the docker containers. 
The hub provides access to \projects, to settings of services and  \reports.
\users{} do all the work in the \notebooks, which run in \containers{} that can be launched and accessed on the hub. Information about system resources and usage of \containers{} can be viewed.
Fig. \ref{fig:kooplex_frontend}. shows the frontpage of \kooplex{}{} with the basic functionalities.

\subsection{Projects}
\projects{} define a storage space and  collaboration between its members. 
A \user{} can create a new, empty \project{} or can clone any of the \projects{} they have access to according to the \project's \scope.
The \user{} that created the \project{} has special attributes and is called the owner of the \project. The \project{} owner and the administrators have the right to configure the \project. 
The environment (any type of packages, modules) are also a property of the project. This way one can ensure that codes developed will run the same way and their work will be reproducible by any other collaborator.
\project{} configuration includes the selection of the (docker) image from which the \container{} is created, options for mounting \volumes, changing of \project's \scope{} and the options to add or remove collaborators.

A new collaboration is born two ways:
\begin{itemize}
    \item either when a \user, who created the \project{}  adds other users to it
    \item or when the user joins an existing public (or internal) \project.
\end{itemize} 

For each \user{} there is a private \emph{workdir} for each of its \project{} (see Section \ref{sec:containers}. for the details of the folder structure).
Collaborators of a \project{} share data in the \project's shared folder and \reports{}{} are automatically accessible for all the collaborators. A \user{} being a member of a \project{} can decide to leave it anytime.

\subsection{Reports}
\label{reports}
A \report{} is a \user's (scientific) product from a given \project.
Static or interactive HTML files provide rich means to visualize and explain the results. \notebooks{} (Jupyter, R Markdown, etc) provide tools to create and present interactive reports or dashboards and even REST services, that actively serves user queries (See for example use case \ref{sec:cdh}).

The \scope{} of a \report{} can be \emph{private}, \emph{internal} or \emph{public}.
Private \reports{} are visible to its creator only, whereas internal \report{} entry points will be shown to all \users{} logged in the system. Public \reports{} are listed publicly under the hub's \emph{report} menu.

Different versions of a \report{} can co-exist and a shortcut to the latest \report{} is provided.
The creator of the \report{} has the right to reconfigure the \scope{} of the \report, put an extra password on it or delete it.

The simplest way to create a \emph{Report} is to convert a \notebook{} into HTML or PDF. \emph{Pyviz.org}, Bokeh\footnote{\url{https://bokeh.pydata.org}} or Dash by Plotly\footnote{\url{https://plot.ly/products/dash/}} provide python packages to create interactive visualization in a Jupyter notebook and convert them into a self-contained document. They can be embedded into other HTML pages as one large object, or as a set of sub-components templated individually.
Bokeh, Jupyter widgets\footnote{\url{https://github.com/jupyter-widgets/ipywidgets}} and scriptedforms\footnote{\url{https://github.com/SimonBiggs/scriptedforms}} allow to create interactive pages, which communicate with a server. This has the advantage that data for the plots can be recalculated and stored on demand and need not to store all the data at once.

A published REST API (e.g. in a Jupyter notebook) is an online service, which provides data, static or dynamic plots and images. This feature allows to embed content into another website with customizable content and without the need of any webdeveloper skills or the help of system administrators.

\section{\kooplexconfig}
\label{sec:config}

\subsection{Virtualization}
\begin{figure}[h!]
 \resizebox {\textwidth} {!} {
 \begin{tikzpicture}
 
   \draw[fill=gray!20] (0ex, 0ex) -- (48ex, 0ex) -- (48ex, -6ex) -- (0ex, -6ex) -- cycle (0ex, -3ex) node[right] {\includegraphics[height=1em]{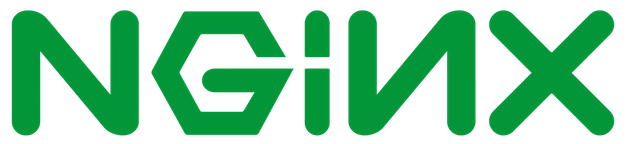}};
   \draw[fill=gray!20] (54ex, 0ex) -- (102ex, 0ex) -- (102ex, -6ex) -- (54ex, -6ex) -- cycle (54ex, -3ex) node[right] {\includegraphics[height=1.3em]{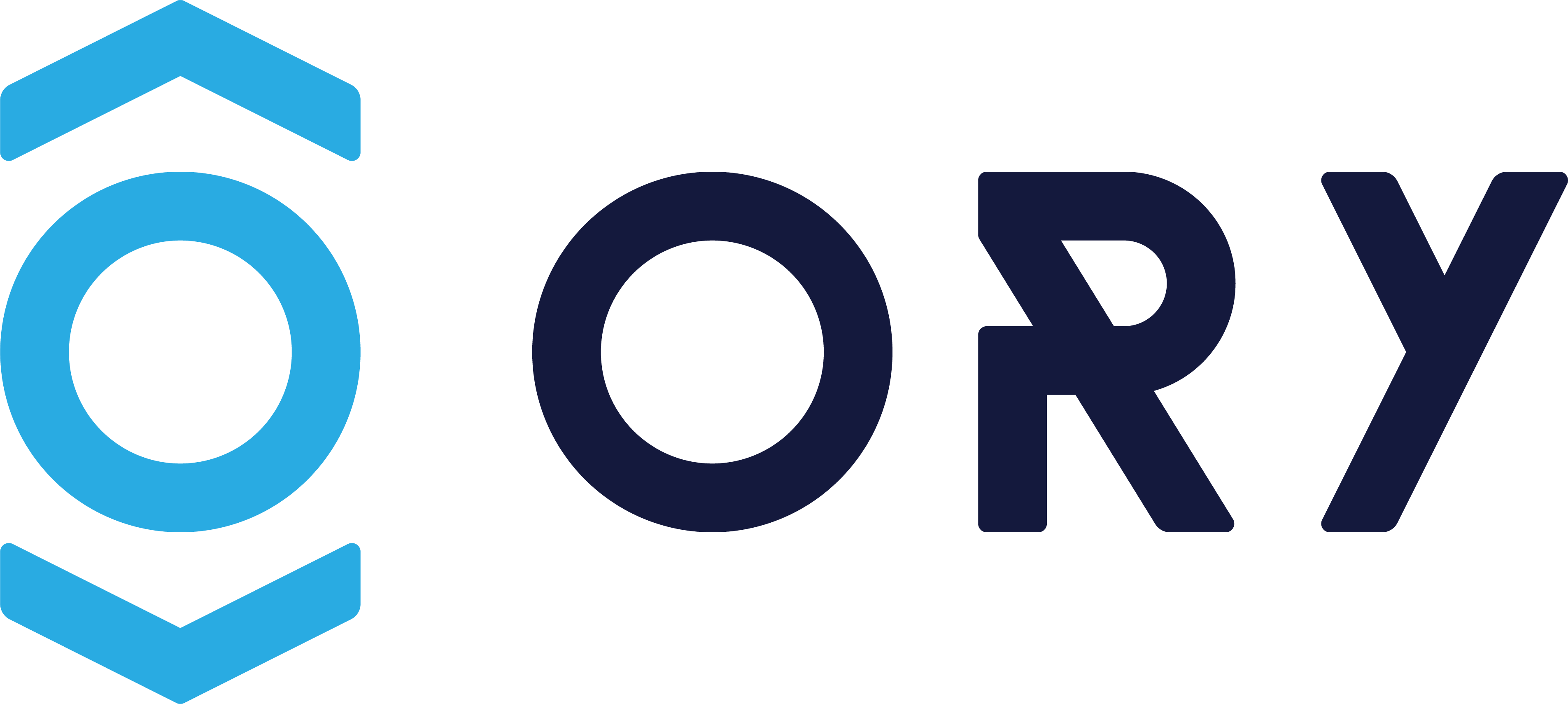}} (62ex, -3.2ex) node[right] {Hydra};
   \draw[fill=gray!20] (0ex, -10ex) -- (43ex, -10ex) -- (43ex, -16ex) -- (0ex, -16ex) -- cycle (0ex, -13ex) node[right] {\includegraphics[height=1em]{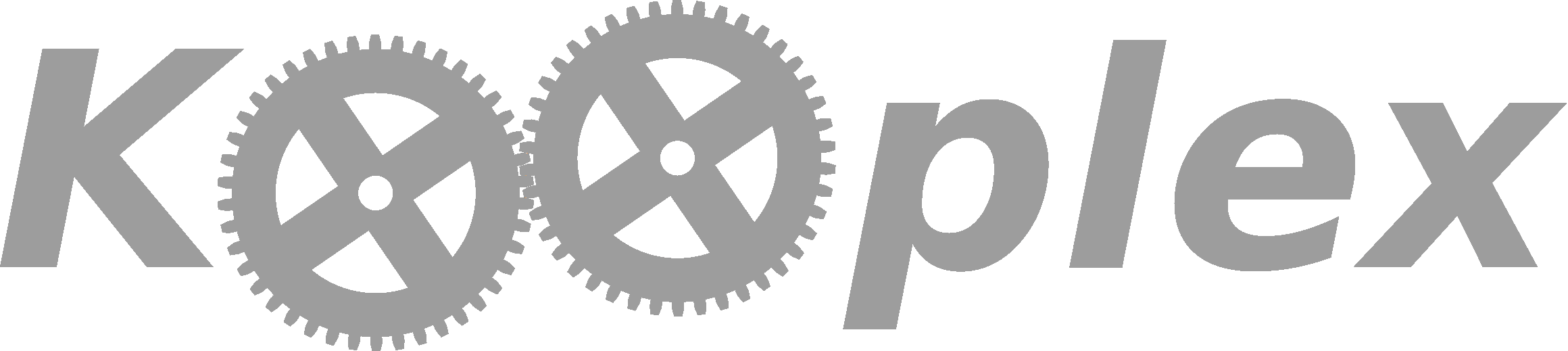} hub};
   \draw[fill=gray!30] (40ex, -15ex) ellipse [x radius = 4ex, y radius = .5ex, start angle = 0, end angle = 360];
   \draw[fill=gray!20] (44ex, -15ex) arc (180:360:-4ex and .5ex) -- +(0, -3ex) arc (180:360:4ex and .6ex) -- +(0, 3ex);
   \draw[fill=gray!20] (0ex, -20ex) -- (12ex, -20ex) -- (12ex, -26ex) -- (0ex, -26ex) -- cycle (0ex, -23ex) node[right] {\small{}Hub DB};
   \draw[fill=gray!30] (10ex, -25ex) ellipse [x radius = 4ex, y radius = .5ex, start angle = 0, end angle = 360];
   \draw[fill=gray!20] (14ex, -25ex) arc (180:360:-4ex and .5ex) -- +(0, -3ex) arc (180:360:4ex and .6ex) -- +(0, 3ex);

   \draw[fill=gray!20] (15ex, -20ex) -- (28ex, -20ex) -- (28ex, -26ex) -- (15ex, -26ex) -- cycle (15ex, -23ex) node[right] {\small{}Ldap};
   \draw[fill=gray!30] (25ex, -25ex) ellipse [x radius = 4ex, y radius = .5ex, start angle = 0, end angle = 360];
   \draw[fill=gray!20] (29ex, -25ex) arc (180:360:-4ex and .5ex) -- +(0, -3ex) arc (180:360:4ex and .6ex) -- +(0, 3ex);

   \draw[fill=gray!20] (30ex, -20ex) -- (43ex, -20ex) -- (43ex, -26ex) -- (30ex, -26ex) -- cycle (29.5ex, -23ex) node[right] {\small{}Monitor};
   \draw[fill=gray!30] (40ex, -25ex) ellipse [x radius = 4ex, y radius = .5ex, start angle = 0, end angle = 360];
   \draw[fill=gray!20] (44ex, -25ex) arc (180:360:-4ex and .5ex) -- +(0, -3ex) arc (180:360:4ex and .6ex) -- +(0, 3ex);

   \draw[fill=gray!20] (45ex, -10ex) -- (60ex, -10ex) -- (60ex, -16ex) -- (58ex, -16ex) -- (58ex, -31ex) -- (93ex, -31ex) -- (93ex, -32ex) -- (5ex, -32ex) -- (5ex, -31ex) -- (57ex, -31ex) -- (57ex, -16ex) -- (45ex, -16ex) -- cycle (45ex, -13ex) node[right] {Proxy};

   \draw[fill=gray!20] (61ex, -10ex) -- (81ex, -10ex) -- (81ex, -16ex) -- (61ex, -16ex) -- cycle (61ex, -13ex) node[right] {Gitea } (74ex, -13ex)  node[left] {\includegraphics[height=2em]{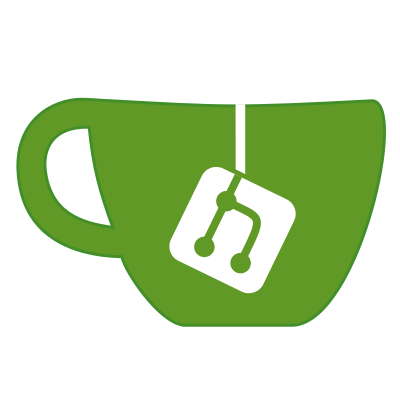}} ;
   \draw[fill=gray!20] (61ex, -20ex) -- (79ex, -20ex) -- (79ex, -26ex) -- (61ex, -26ex) -- cycle (61ex, -23ex) node[right] {\small{}Gitea DB};
   \draw[fill=gray!30] (77ex, -25ex) ellipse [x radius = 4ex, y radius = .5ex, start angle = 0, end angle = 360];
   \draw[fill=gray!20] (81ex, -25ex) arc (180:360:-4ex and .5ex) -- +(0, -3ex) arc (180:360:4ex and .6ex) -- +(0, 3ex);

   \draw[fill=gray!20] (82ex, -10ex) -- (102ex, -10ex) -- (102ex, -16ex) -- (82ex, -16ex) -- cycle (82ex, -13ex) node[right] {Seafile} (95ex, -13ex)  node[left] {\includegraphics[height=2em]{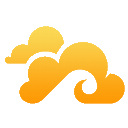}};
   \draw[fill=gray!20] (82ex, -20ex) -- (100ex, -20ex) -- (100ex, -26ex) -- (82ex, -26ex) -- cycle (82ex, -23ex) node[right] {\small{}Seafile DB};
   \draw[fill=gray!30] (98ex, -25ex) ellipse [x radius = 4ex, y radius = .5ex, start angle = 0, end angle = 360];
   \draw[fill=gray!20] (102ex, -25ex) arc (180:360:-4ex and .5ex) -- +(0, -3ex) arc (180:360:4ex and .6ex) -- +(0, 3ex);

   \draw[fill=gray!20] (5ex, -35ex) -- (18ex, -35ex) -- (18ex, -41ex) -- (5ex, -41ex) -- cycle (5ex, -38ex) node[right] {\small{}Notebook$_1$};
   \draw[fill=gray!30] (15ex, -40ex) ellipse [x radius = 4ex, y radius = .5ex, start angle = 0, end angle = 360];
   \draw[fill=gray!20] (19ex, -40ex) arc (180:360:-4ex and .5ex) -- +(0, -3ex) arc (180:360:4ex and .6ex) -- +(0, 3ex);

   \draw[fill=gray!70] (20ex, -35ex) -- (33ex, -35ex) -- (33ex, -41ex) -- (20ex, -41ex) -- cycle (20ex, -38ex) node[right] {\small{}Notebook$_2$};
   \draw[fill=gray!80] (30ex, -40ex) ellipse [x radius = 4ex, y radius = .5ex, start angle = 0, end angle = 360];
   \draw[fill=gray!70] (34ex, -40ex) arc (180:360:-4ex and .5ex) -- +(0, -3ex) arc (180:360:4ex and .6ex) -- +(0, 3ex);

   \draw[fill=gray!20] (40ex, -35ex) -- (53ex, -35ex) -- (53ex, -41ex) -- (40ex, -41ex) -- cycle (40ex, -38ex) node[right] {\small{}Notebook$_p$};
   \draw[fill=gray!30] (50ex, -40ex) ellipse [x radius = 4ex, y radius = .5ex, start angle = 0, end angle = 360];
   \draw[fill=gray!20] (54ex, -40ex) arc (180:360:-4ex and .5ex) -- +(0, -3ex) arc (180:360:4ex and .6ex) -- +(0, 3ex);

   \draw[fill=gray!20] (60ex, -35ex) -- (73ex, -35ex) -- (73ex, -41ex) -- (60ex, -41ex) -- cycle (60ex, -38ex) node[right] {\small{}Report$_1$};
   \draw[fill=gray!30] (70ex, -40ex) ellipse [x radius = 4ex, y radius = .5ex, start angle = 0, end angle = 360];
   \draw[fill=gray!20] (74ex, -40ex) arc (180:360:-4ex and .5ex) -- +(0, -3ex) arc (180:360:4ex and .6ex) -- +(0, 3ex);

   \draw[fill=gray!20] (80ex, -35ex) -- (93ex, -35ex) -- (93ex, -41ex) -- (80ex, -41ex) -- cycle (80ex, -38ex) node[right] {\small{}Report$_q$};
   \draw[fill=gray!30] (90ex, -40ex) ellipse [x radius = 4ex, y radius = .5ex, start angle = 0, end angle = 360];
   \draw[fill=gray!20] (94ex, -40ex) arc (180:360:-4ex and .5ex) -- +(0, -3ex) arc (180:360:4ex and .6ex) -- +(0, 3ex);
 \end{tikzpicture}
 }
  \caption{The \kooplex \hspace{1pt} architecture. All the components are dockerized. In the figure the rectangles represent docker containers and the cylinders stand for docker volumes. Nginx is the web server, that directs traffic between the \user \hspace{1pt} and the platform with static routing. Hydra manages user authentication. The configurable-http-proxy is a dynamic routing system that connects the \user \hspace{1pt} to the \notebooks .
  Ldap acts as a local usermanager and provides the \containers \hspace{1pt} with the right user ids. \reports \hspace{1pt} are served from \containers \hspace{1pt} as well. \notebooks \hspace{1pt} can have further access to compute nodes via a job scheduler system.
}
  \label{fig:architecture}
\end{figure}
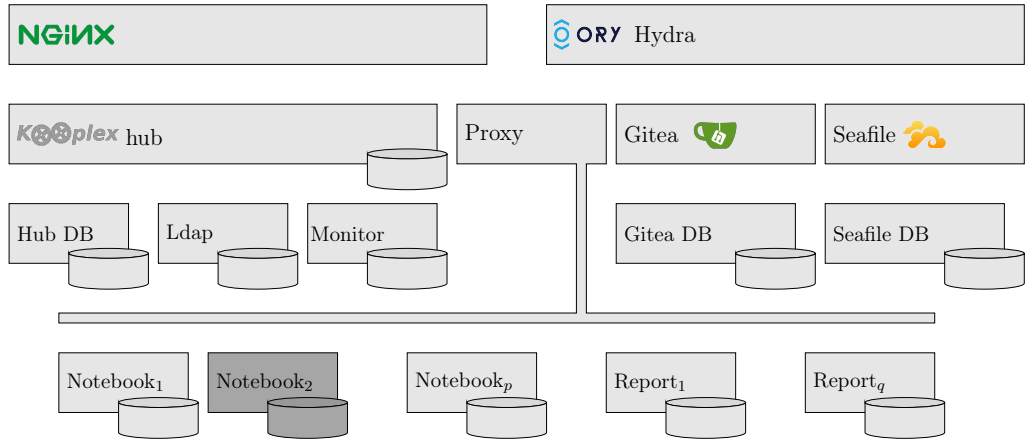

Docker engine\cite{Merkel:2014:DLL:2600239.2600241} provides the skeleton of the system, which is a light\-weight virtualization or containerization solution. Following modularity principles all the components of \kooplex{} platform run in a separate Docker container.
While Docker isolates all of the software components, it eases their orchestrated installation with \emph{docker-compose} in exchange for a negligible amount of system resources.
Fig. \ref{fig:architecture} shows the architecture of the \kooplex{} platform and how each component is related to the other. 
\kooplex{} consists of five components: the \kooplexhub{} (webpage, auxiliary services, local user database etc.), the \notebooks{} (Jupyter, Rstudio etc.), \emph{service providers} (Gitea, GitLab, OwnCloud, SeaFile, etc), \emph{storages} (data or functional volumes) and optionally the \emph{compute nodes}. 
Each of these components might consist of several docker containers (e.g. the service itself and its database) and their internal communication happens within a separated network.

One of our motivation with the \kooplex{} platform was to create a mechanism, which allows to run the calculations (the \notebooks{}) close to the data. 
In order to scale up and distribute computational tasks we have investigated several state of the art solutions. According to our experience Kubernetes\footnote{\url{https://kubernetes.io/}} is a well-configurable and flexible automated deployment system, that extends Docker functionalities. With Kubernetes it is possible to spawn \containers{} to the host closest to the data while considering the machines' workload.


\subsection{Hydra and user authenticaton}
\label{sec:hydra}
Django is capable of administering of users and communicate with service providers but for the sake of modularity and to keep the platform robust we use an independent authentication layer, Hydra\footnote{https://github.com/ory/hydra}, to keep the user database separate from \kooplex. This motivates the usage of already existing user databases, which keeps the management simpler and prevents \kooplex{} to store any \user{} provided password.

\subsection{Local user database, LDAP}
Regardless of the authentication method into a \kooplex{} instance it is necessary to have a local copy of the user database, which is stored in LDAP \footnote{\url{https://www.openldap.org/}} . This ensures the unequivocal user identification through all \notebooks{} and services. 

\subsection{Volumes}
\subsubsection*{Storage volumes}
\label{sec:volumes}
Docker facilitates the usage of data storages. Once a docker volume is created it can be mounted to any \container. This is especially useful when we want to access data from physically different storage systems. Within \kooplex{} access to these storages can be restricted to \users{} or groups of \users{}, which is necessary in case of datahubs with sensitive content.


\subsubsection*{Functional volumes}
Sharing of environments between \users{} is possible on multiple levels. One way is by using \functionalvolumes{} They are independent objects, don't belong to any project and can be attached to any \container. \users{} might place here sample data and other files that are not meant to be version controlled.
\functionalvolumes{} may contain different set of packages, samples of datasets, test files etc.

\subsection{Containers and folder structure}
\label{sec:containers}
A \user{} can create \containers. In a \container{} any number of \projects{} can be accessed, and each of them will appear as a separate folder (as shown in Fig. \ref{fig:filesystemlayout}.).
\containers{} are spawned from Docker images. In each \container{} a browser based service, a \notebook{} runs such as Jupyter notebook server or R Studio server etc.

In Table \ref{table:folderspersistent}. and \ref{table:foldersservice}. we listed the persistent directories that \kooplex{} prepares for each \user, which can be accessed in a \container{}. There is a list of folders, which belong to the attached \service{} (See Table \ref{table:foldersservice}. for further description). 
In Fig.~\ref{fig:filesystemlayout}. we illustrate the folder structure of a  \container.
 \usetikzlibrary{decorations.pathreplacing}

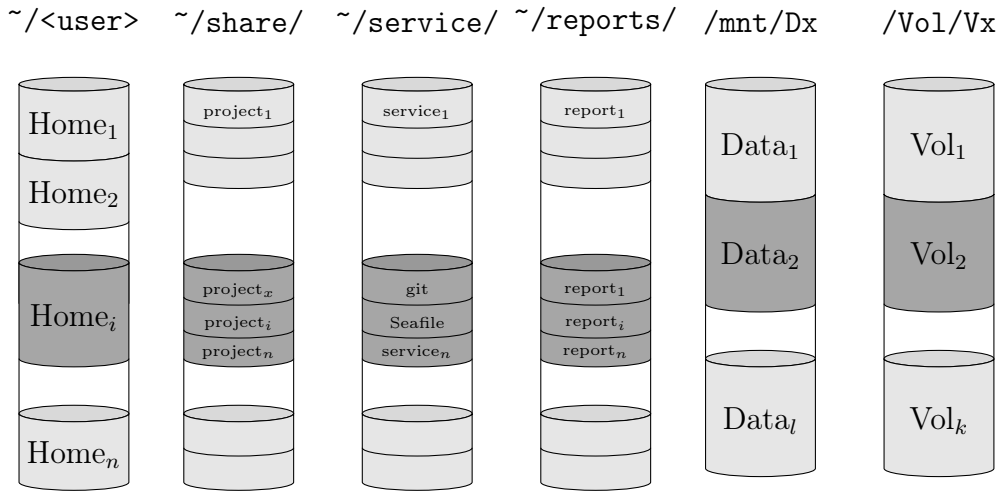
\begin{figure}[h!]
 \resizebox {\textwidth} {!} {
 \begin{tikzpicture}
   
   \draw (12ex, 2em) node {\url{~/<user>}};
   \draw[fill=gray!30] (12ex, 0) ellipse [x radius = 4ex, y radius = .5ex, start angle = 0, end angle = 360];
   \draw[fill=gray!20] (16ex, 0) arc (180:360:-4ex and .5ex) -- +(0, -5ex) arc (180:360:4ex and .6ex) -- +(0, 5ex);
   \draw (12ex, -3ex) node {Home$_1$};
   \draw[fill=gray!20] (16ex, -5ex) arc (180:360:-4ex and .6ex) -- +(0, -5ex) arc (180:360:4ex and .6ex) -- +(0, 5ex);
   \draw (12ex, -8ex) node {Home$_2$};
   \draw (16ex, -10ex) -- +(0, -6ex);
   \draw (8ex, -10ex) -- +(0, -6ex);
   \draw[fill=gray!80] (12ex, -13ex) ellipse [x radius = 4ex, y radius = .6ex, start angle = 0, end angle = 360];
   \draw[fill=gray!70] (16ex, -13ex) arc (180:360:-4ex and .6ex) -- +(0, -7ex) arc (180:360:4ex and .6ex) -- +(0, 7ex);
   \draw (12ex, -17ex) node {Home$_i$};
   \draw (16ex, -20ex) -- +(0, -4ex);
   \draw (8ex, -20ex) -- +(0, -4ex);
   \draw[fill=gray!30] (12ex, -24ex) ellipse [x radius = 4ex, y radius = .6ex, start angle = 0, end angle = 360];
   \draw[fill=gray!20] (16ex, -24ex) arc (180:360:-4ex and .6ex) -- +(0, -5ex) arc (180:360:4ex and .6ex) -- +(0, 5ex);
   \draw (12ex, -27ex) node {Home$_n$};

   \draw (24ex, 2em) node {\url{~/share/}};
   \draw[fill=gray!30] (24ex, 0) ellipse [x radius = 4ex, y radius = .5ex, start angle = 0, end angle = 360];
   \draw[fill=gray!20] (28ex, 0) arc (180:360:-4ex and .5ex) -- +(0, -7ex) arc (180:360:4ex and .6ex) -- +(0, 7ex);
   \draw (24ex, -2ex) node {\tiny{}project$_1$};
   \draw (20ex, -2.6ex) arc (180:360:4ex and .6ex);
   \draw (20ex, -4.8ex) arc (180:360:4ex and .6ex);
   \draw (28ex, -7ex) -- +(0, -6ex);
   \draw (20ex, -7ex) -- +(0, -6ex);
   \draw[fill=gray!80] (24ex, -13ex) ellipse [x radius = 4ex, y radius = .6ex, start angle = 0, end angle = 360];
   \draw[fill=gray!70] (28ex, -13ex) arc (180:360:-4ex and .6ex) -- +(0, -7ex) arc (180:360:4ex and .6ex) -- +(0, 7ex);
   \draw (24ex, -15ex) node {\tiny{}project$_x$};
   \draw (24ex, -17.4ex) node {\tiny{}project$_i$};
   \draw (24ex, -19.5ex) node {\tiny{}project$_n$};
   \draw (20ex, -15.5ex) arc (180:360:4ex and .6ex);
   \draw (20ex, -17.9ex) arc (180:360:4ex and .6ex);
   \draw (28ex, -20ex) -- +(0, -4ex);
   \draw (20ex, -20ex) -- +(0, -4ex);
   \draw[fill=gray!30] (24ex, -24ex) ellipse [x radius = 4ex, y radius = .6ex, start angle = 0, end angle = 360];
   \draw[fill=gray!20] (28ex, -24ex) arc (180:360:-4ex and .6ex) -- +(0, -5ex) arc (180:360:4ex and .6ex) -- +(0, 5ex);
   \draw (20ex, -26.5ex) arc (180:360:4ex and .6ex);

  \draw (37ex, 2em) node {\url{~/service/}};
  \draw[fill=gray!30] (37ex, 0) ellipse [x radius = 4ex, y radius = .5ex, start angle = 0, end angle = 360];
  \draw[fill=gray!20] (41ex, 0) arc (180:360:-4ex and .5ex) -- +(0, -7ex) arc (180:360:4ex and .6ex) -- +(0, 7ex);
   \draw (37ex, -2ex) node {\tiny{}service$_1$};
   \draw (33ex, -2.6ex) arc (180:360:4ex and .6ex);
   \draw (33ex, -4.8ex) arc (180:360:4ex and .6ex);
   \draw (41ex, -7ex) -- +(0, -6ex);
   \draw (33ex, -7ex) -- +(0, -6ex);
   \draw[fill=gray!80] (37ex, -13ex) ellipse [x radius = 4ex, y radius = .6ex, start angle = 0, end angle = 360];
   \draw[fill=gray!70] (41ex, -13ex) arc (180:360:-4ex and .6ex) -- +(0, -7ex) arc (180:360:4ex and .6ex) -- +(0, 7ex);
   \draw (37ex, -15ex) node {\tiny{}git};
   \draw (37ex, -17.4ex) node {\tiny{}Seafile};
   \draw (37ex, -19.5ex) node {\tiny{}service$_n$};
   \draw (33ex, -15.5ex) arc (180:360:4ex and .6ex);
   \draw (33ex, -17.9ex) arc (180:360:4ex and .6ex);
   \draw (41ex, -20ex) -- +(0, -4ex);
   \draw (33ex, -20ex) -- +(0, -4ex);
   \draw[fill=gray!30] (37ex, -24ex) ellipse [x radius = 4ex, y radius = .6ex, start angle = 0, end angle = 360];
   \draw[fill=gray!20] (41ex, -24ex) arc (180:360:-4ex and .6ex) -- +(0, -5ex) arc (180:360:4ex and .6ex) -- +(0, 5ex);
   \draw (33ex, -26.5ex) arc (180:360:4ex and .6ex);

  \draw (50ex, 2em) node {\url{~/reports/}};
  \draw[fill=gray!30] (50ex, 0) ellipse [x radius = 4ex, y radius = .5ex, start angle = 0, end angle = 360];
  \draw[fill=gray!20] (54ex, 0) arc (180:360:-4ex and .5ex) -- +(0, -7ex) arc (180:360:4ex and .6ex) -- +(0, 7ex);
   \draw (50ex, -2ex) node {\tiny{}report$_1$};
   \draw (46ex, -2.6ex) arc (180:360:4ex and .6ex);
   \draw (46ex, -4.8ex) arc (180:360:4ex and .6ex);
   \draw (54ex, -7ex) -- +(0, -6ex);
   \draw (46ex, -7ex) -- +(0, -6ex);
   \draw[fill=gray!80] (50ex, -13ex) ellipse [x radius = 4ex, y radius = .6ex, start angle = 0, end angle = 360];
   \draw[fill=gray!70] (54ex, -13ex) arc (180:360:-4ex and .6ex) -- +(0, -7ex) arc (180:360:4ex and .6ex) -- +(0, 7ex);
   \draw (50ex, -15ex) node {\tiny{}report$_1$};
   \draw (50ex, -17.4ex) node {\tiny{}report$_i$};
   \draw (50ex, -19.5ex) node {\tiny{}report$_n$};
   \draw (46ex, -15.5ex) arc (180:360:4ex and .6ex);
   \draw (46ex, -17.9ex) arc (180:360:4ex and .6ex);
   \draw (54ex, -20ex) -- +(0, -4ex);
   \draw (46ex, -20ex) -- +(0, -4ex);
   \draw[fill=gray!30] (50ex, -24ex) ellipse [x radius = 4ex, y radius = .6ex, start angle = 0, end angle = 360];
   \draw[fill=gray!20] (54ex, -24ex) arc (180:360:-4ex and .6ex) -- +(0, -5ex) arc (180:360:4ex and .6ex) -- +(0, 5ex);
   \draw (46ex, -26.5ex) arc (180:360:4ex and .6ex);


   \draw (62ex, 2em) node {\url{/mnt/Dx}};
   \draw[fill=gray!30] (62ex, 0) ellipse [x radius = 4ex, y radius = .5ex, start angle = 0, end angle = 360];
   \draw[fill=gray!20] (66ex, 0) arc (180:360:-4ex and .5ex) -- +(0, -8ex) arc (180:360:4ex and .6ex) -- +(0, 8ex);
   \draw (62ex, -4.5ex) node {Data$_1$};
   \draw[fill=gray!70] (66ex, -8ex) arc (180:360:-4ex and .6ex) -- +(0, -8ex) arc (180:360:4ex and .6ex) -- +(0, 8ex);
   \draw (62ex, -12.5ex) node {Data$_2$};
   \draw (66ex, -16ex) -- +(0, -4ex);
   \draw (58ex, -16ex) -- +(0, -4ex);
   \draw[fill=gray!30] (62ex, -20ex) ellipse [x radius = 4ex, y radius = .6ex, start angle = 0, end angle = 360];
   \draw[fill=gray!20] (66ex, -20ex) arc (180:360:-4ex and .6ex) -- +(0, -8ex) arc (180:360:4ex and .6ex) -- +(0, 8ex);
   \draw (62ex, -24.5ex) node {Data$_l$};
   
   \draw (75ex, 2em) node {\url{/Vol/Vx}};
   \draw[fill=gray!30] (75ex, 0) ellipse [x radius = 4ex, y radius = .5ex, start angle = 0, end angle = 360];
   \draw[fill=gray!20] (79ex, 0) arc (180:360:-4ex and .5ex) -- +(0, -8ex) arc (180:360:4ex and .6ex) -- +(0, 8ex);
   \draw (75ex, -4.5ex) node {Vol$_1$};
   \draw[fill=gray!70] (79ex, -8ex) arc (180:360:-4ex and .6ex) -- +(0, -8ex) arc (180:360:4ex and .6ex) -- +(0, 8ex);
   \draw (75ex, -12.5ex) node {Vol$_2$};
   \draw (79ex, -16ex) -- +(0, -4ex);
   \draw (71ex, -16ex) -- +(0, -4ex);
   \draw[fill=gray!30] (75ex, -20ex) ellipse [x radius = 4ex, y radius = .6ex, start angle = 0, end angle = 360];
   \draw[fill=gray!20] (79ex, -20ex) arc (180:360:-4ex and .6ex) -- +(0, -8ex) arc (180:360:4ex and .6ex) -- +(0, 8ex);
   \draw (75ex, -24.5ex) node {Vol$_k$};





 \end{tikzpicture}
 }
  \caption{The filesystem view. Each user and project have their own persistent storage. There are parts that overlap or span across a collaboration. On the figure the areas filled with darker color illustrate the directories that a particular \user{}  have access to in a \container. Every \user{}  has a private folder. In the \emph{share} directory the shared space of each connected project appear. A \emph{service} is a component such as version control system, cloud based filesharing etc, of which content can be accessed from the  \container. \emph{Reports} are published notebooks, of which snapshots are stored in read-only folders. $Data_l$ and $Vol_k$  refer to  \storagevolumes and \functionalvolumes{}. }
  \label{fig:filesystemlayout}
\end{figure}

\begin{table}[h!]
\caption{Persistent folders in each\container}
    \begin{tabular}{p{2.5cm} p{10cm}}
    \addlinespace[0.1cm]
        \hline
        Folder &  Description  \\
        \hline  \hline
         \addlinespace[0.2cm]
         
        \textbf{\home} & is completely private \\[.5em]
        
        \textbf{\workdir} & is a private working directory for each project \\[.5em]
        
        \textbf{\share} & folder is shared between all the project members with read/write access depending on their role \\[.5em]

        \textbf{\functionalvolumes} & store additional packages, softwares, they are writable only for the maintainer \\[.5em]

        \textbf{\storagevolumes} & access for the large datasets \\[.5em]

        \textbf{\report} & folder is for preparing the publishable reports or applications and it also stores the different versions of previously committed reports 
    \label{table:folderspersistent}
    \end{tabular}
\end{table}

\begin{table}
\caption{Folder structure for \services{} in each \container}
    \begin{tabular}{p{2.5cm} p{10cm}}
    
    \addlinespace[0.1cm]
        \hline
        Folder &  Description  \\
        \hline  \hline
         \addlinespace[0.2cm]
        \textbf{\seafile} & files stored in this directory are automatically synchronized with the attached cloud like service (Seafile, OwnCloud, NextCloud etc). These platforms offer the possibility to share files and folders independently from \kooplex, and have desktop clients for private synchronization as well.  \\[.5em]
        
        \textbf{\gitfolder} & in each subdirectory a the version controlled content is to be found that reside in one of the repositories (Github, GitLab, Gitea etc.) to which the \user{} provided credentials (these are tokens and public keys and not passwords)\\[.5em]
        
     \label{table:foldersservice}
    \end{tabular}
\end{table}

 \subsection{Customizing the working environment}
 Customization of the  environment is important for explorative work and the mechanism of \user{} defined persistent environments accompanying the notebooks help to create reproducible workflows. In \kooplex{} a \user{} can create \functionalvolumes{} to store additional pieces of code, or small size data files to pass information across various environments or support project independent reproducibility. It is also possible to create a cross platform representation of the environment as an encapsulated Singularity\footnote{\url{https://github.com/sylabs/singularity}} image. These images can be published under various \scopes{} just like \reports.

 \subsection{Services}
 Services such as Gitea\footnote{\url{https://gitea.io}}, Seafile\footnote{\url{https://www.seafile.com}} etc. add their own functionalities and extra content to the \containers.
 The more such services are available on the platform instance, the higher is the chance that someone finds motivation to join in and start to collaborate. User level settings of these \services{} are accessible from the \kooplexhub's frontend.

\subsection{Compute nodes}
Computationally intensive jobs can be submitted to separate computers or docker containers with dedicated resources. There are several softwares to manage job queue system, such as SLURM\footnote{\url{https://slurm.schedmd.com/}}, PBS\footnote{\url{https://github.com/pbspro/pbspro}} or HTCondor\footnote{\url{http://research.cs.wisc.edu/htcondor/}}. The client side softwares of these \services{} are compiled into the images and thus available in the \containers.

\section{A brief overview of other platforms}
\label{sec:platform-comparison}

In this section we give an overview of currently available not necessarily complete list of data science and business intelligence platforms. 
It helps the reader to put \kooplex{} into context. 
We enumerate and evaluate some key aspects (See Table \ref{table:asp1}. and \ref{table:asp2}.) that one might take into account when choosing the suitable solution from the plethora of platform implementations.

\newcommand{\contwork}{(workflow)}
\newcommand{\repr}{(reproducibility)}
\newcommand{\repo}{(reporting)}
\newcommand{\collab}{(collaboration)}
\newcommand{\open}{(open source)}
\newcommand{\userm}{(user management)}
\newcommand{\scale}{(scaling)}
\newcommand{\env}{(customization)}

\newcommand{\econtwork}{(\emph{workflow})}
\newcommand{\erepr}{(\emph{reproducibility})}
\newcommand{\erepo}{(\emph{reporting})}
\newcommand{\ecollab}{(\emph{collaboration})}
\newcommand{\eopen}{(\emph{open source})}
\newcommand{\euserm}{(\emph{user management})}
\newcommand{\escale}{(\emph{scaling})}
\newcommand{\eenv}{(\emph{customization})} 

\begin{table}[h!]
\caption{General key aspects of VREs}
    \begin{tabular}{p{3cm} p{9.5cm}}
    \addlinespace[0.1cm]
        \hline
        Term &  Description  \\
        \hline  \hline
         \addlinespace[0.2cm]
         
        \textbf{Complete workflow} & the possibility to conduct the whole data processing, calculation and visualization from the beginning till the end on the same platform with possibly applying functionalities via external tools or services. \\[.5em]

        \textbf{Customizable environment} & a modular server structure with user rights granted to customize the working environment. \\[.5em]

        \textbf{Scalability} & performance tuning by adding extra resources at any time. \\[.5em]

        \textbf{Flexible user management} & authentication to the VRE can also be done via external identity providers ( See Section \ref{sec:hydra}.) \\[.5em]

        \textbf{Open source} & source code open for inspection and with permission granted to freely re-use and/or redistribute.\footnote{\url{https://opensource.org/osd}} \\[.5em]

        \textbf{Browser based} & the \user{} does not need to install additional programmes to use the platform. \\[.5em]
        \label{table:asp1}
    \end{tabular}
\end{table}

\begin{table}
\begin{center}
\textbf{Comparison of platforms I.} \\

\begin{tabular}{l|c|c|c|c|c|c|c|c|c}
\addlinespace[0.1cm]
\hline
VRE &\contwork &\env &\scale & \open  \\
\hline \hline
\addlinespace[0.2cm]
Colab8      & \chm & \textbf{L} & \textbf{P} &  \textbf{x} \\
CodeOcean  & \chm & \chm & \textbf{P} & \textbf{x} \\
JEODPP     & \chm & \chm & \textbf{A} &  \textbf{x} \\
WholeTale  & \textbf{L} & \textbf{L} & \textbf{x} &   \textbf{x} \\
SciServer  & \chm & \chm & \chm & \textbf{x} \\
Crewspark  & \chm & \textbf{L} & \textbf{NI}  &  \chm \\
Anaconda*   & \textbf{P} & \textbf{P} & \textbf{P} &  \textbf{P} \\
JupyterHub* & \chm & \chm & \chm & \chm \\
CoCalc*     & \textbf{L} & \textbf{L} & \textbf{L} &   \chm \\
\kooplex{}*   & \chm & \chm & \chm &  \chm\\

\end{tabular}
\caption{\label{table:comp1} General comparison of VREs. \textbf{*}: needs to be installed, \textbf{L}: this feature is available only in limited form, \textbf{P}: paid service, \textbf{A}: for academic use only, \textbf{NI}: we have no information about it and \textbf{x}: this feature is 'not present' in the platform.  }
\end{center}
\end{table}

\begin{table}
\caption{Collaborative key aspects of VREs}
    \begin{tabular}{p{3.4cm} p{9.4cm}}
    \addlinespace[0.1cm]
        \hline
        Term &  Description  \\
        \hline   \hline
        \addlinespace[0.2cm]
        
        \textbf{Collaboration} & various methods implements sharing of data and/or code associated with \users{} (See Section \ref{sec:collab}.) \\ & \\[.5em]
    
        \textbf{Reporting} & an output of work to be presented to a certain audience (See Section \ref{sec:report}.) \\ & \\[.5em]
   
        \textbf{Reproducibility} & data and code compiled with and bound to the report to support its \emph{reproducibility} and to prove \emph{credibility}\footnote{\url{https://osf.io/s9tya}} (See Section \ref{sec:repr}.) \\[.5em]
    \label{table:asp2}
    \end{tabular}
\end{table}

\begin{table}
\begin{center}
\textbf{Comparison of platforms II.}

\begin{tabular}{c|c|c| c| c}
\addlinespace[0.1cm]
\hline
VRE  & \collab &  \repr &\repo & \userm  \\
\hline \hline
Colab     & \textbf{L} & \textbf{x} & \textbf{x} & \textbf{x}\\
CodeOcean &\chm &  \chm & \chm & \textbf{x}\\
JEODPP    & \textbf{x} & \chm & \textbf{x} & \textbf{x}\\
WholeTale & \textbf{L} &\chm & \chm & \textbf{x}\\
SciServer & \textbf{x} & \textbf{x} & \textbf{x} & \textbf{x} \\
Crewspark & \chm & \chm & \chm & \textbf{x}\\
Anaconda  & \textbf{P} &  \textbf{P} & \textbf{P} & \textbf{P}\\
JupyterHub & \textbf{x} & \textbf{x} & \textbf{x} & \chm\\
CoCalc    & \textbf{L} & \textbf{x} & \chm & \textbf{L}\\
\kooplex  & \chm & \chm & \chm &  \chm\\
\end{tabular}
\caption{\label{table:comp2} \textbf{L}: this feature is available only in limited form, \textbf{P}: paid service, \textbf{A}: for academic use only, \textbf{NI}: we have no information about it and \textbf{x}: this feature is 'not present' in the platform.  }
\end{center}
\end{table}

In Table \ref{table:comp1}. and \ref{table:comp2}. we compare different platforms from an individual user or administrator point of view and take into account collaborative aspects.

\subsection{Collaboration}
\label{sec:collab}
In a complex research project, where many disciplines and participants with different skills come together collaboration is a key aspect. It can be manifested on many levels:
        \begin{itemize}
            \item \emph{consecutive:}  a collaborator's input depends on another's output.
            \item \emph{parallel/same time:} collaborators work on the same problem editing the same documents, codes or work with the same data at the same time.
            \item \emph{randomly induced:} individuals randomly join and contribute to each others work. Similar to the case of open source code development.
        \end{itemize} 

\subsection{Reporting}
\label{sec:report}
 It is a functional feature of the platform that \reports{} can be created, which connects the \users{} (researchers, developers) and the public. For instance laymans, funding agencies, companies can be interested in the reported information and can utilise it in their own business or academic work.
 
\subsection{Reproducibility}
\label{sec:repr}
In science reproducibility is an inherent key concept. A study is regarded reproducible if the given specific set of computational functions and chain of analysis (usually specified in terms of code) reproduces exactly the same figures. On the long run scientific progress relies on reproducibility and reuse of research \cite{green2018computational}. 

By the time when some of the relevant questions are answered research goes through different stages and passes certain checkpoints. At these stages written notes, notebooks, test data, codes and figures are created. These collections should be accessible to those who intend to check and reproduce the results starting from step one with the raw data.

\subsubsection{Platforms and repositories}
In this section we focus on a few popular platforms based on the previously detailed notions.

The aspects of reproducibility \erepo{} are already realized as "tales" on \emph{Wholetale.org}  \cite{BRINCKMAN2018}, as "capsules" on \emph{codeocean.com} \cite{boettiger2015introduction} or as "kernels" in \emph{Kaggle}\footnote{\url{http://www.kaggle.com/}}. \emph{Wholetale.org} and \emph{Kaggle} are both \emph{browser based} platforms that employ Jupyter notebook in their VRE, and aim to support users in all phases of the research lifecyle, from data acquisition to publication, including the cross-linking of data, software, workflows and manuscripts. Similarly, in \kooplex{} the whole procedure can be saved either as a \report{} or compiled in a Singularity image.


A versatile solution that incorporates more of the above aspects is \emph{Codeocean}\footnote{\url{https://codeocean.com/}} \cite{boettiger2015introduction} with customizable environment \eenv{} and parallel collaboration \ecollab.  

 \emph{Anaconda}\footnote{\url{https://www.anaconda.com/enterprise}} is a commercial service and its pricing depends either on usage of the computational resources or on the number of users and enabled services. This platform also provides tools for building workflows \econtwork.

The \emph{Sciserver}\footnote{\url{http://www.sciserver.org/}} is a platform intended for academic research and provides free access to many datahubs (SDSS DAS, recount2, JHTDB), large computational resources and offers some collaborative features. 

\emph{Google's Colab}\footnote{\url{https://colab.research.google.com}} offers parallel collaboration \ecollab, \econtwork{} and \euserm{} is automatically fulfilled through Google services. \eenv{},  \escale{}  and \erepr{} are limited. 

An example for data-intensive geospatial computing means is the \emph{JRC Earth Observation Data and Processing Platform} \cite{SOILLE201830,de_marchi_davide_2019_3239239}. In this implementation \econtwork, \escale, \ecollab, \erepo, and \erepr{} are fulfilled.

Strictly speaking the \emph{Experiment Factory} \cite{sochat2018experiment} is not a platform rather it is a repository. It is designed for storing reproducible experiments in the field of behavioral research \erepr{} using docker images. The spawned containers can be downloaded and used locally.

\section{Use cases of \kooplex}
\label{sec:kooplex-usecase}

Installation of \kooplex{} with the default parameters requires the basic knowledge of bash and python commands in linux. A \kooplex{} instance served by an average desktop computer can support the work of a fairly large group, for up to 10--20 users in parallel without they doing computationally intensive calculations.
For a larger userbase it is recommended to spawn the \containers{} to a separate server or to a cluster of servers.

\subsection{Remote Datahubs}
\label{sec:cdh}
\begin{figure}[h!]
    \centering
    \includegraphics[width=400px]{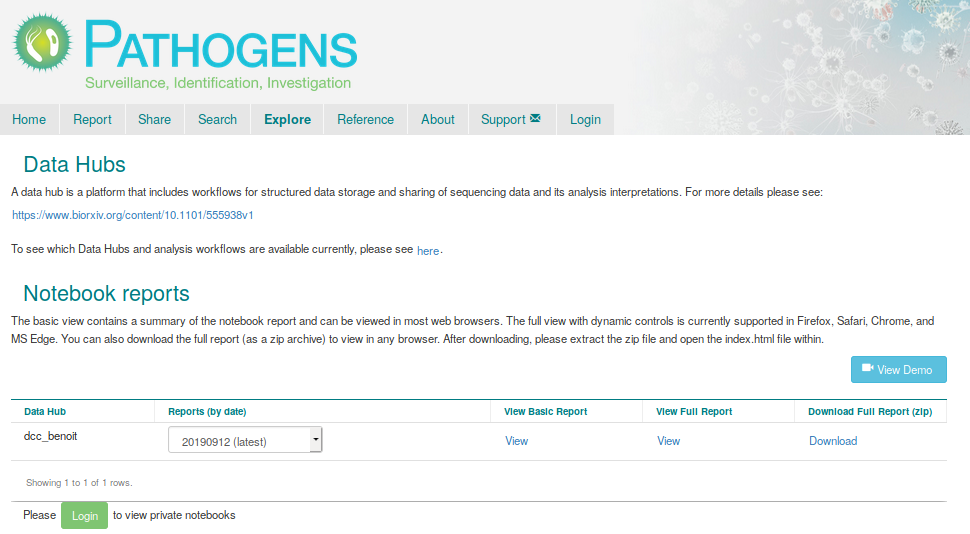}
    \caption{The Pathogens portal at EMBL-EBI. A \kooplex{} portal is running close to the datahubs and one the codes developed in \notebooks{} generates a public report daily}
    \label{fig:kooplex_pathogens}
\end{figure}

We currently operate several \kooplex{} instances for different target user groups. One of the instances is deployed at EMBL-EBI\cite{amid2019compare} for the ENA internally, where users carry out the final steps of the analysis of the residing data and generate visualizations. There are special pieces of code developed in \notebooks{}, from which the automatically generated \reports{} are published and served through the Pathogens portal\footnote{\url{https://www.ebi.ac.uk/ena/pathogens/explore}} (See Fig. \ref{fig:kooplex_pathogens}). 
Note, the two types of \reports{} used most often in this \kooplex{} instance to be rendered in a web browser. The simpler kind is a static HTML that can give a quick overview about the content of the datahub on the given day. Additionally, the interactive report type allow users to filter, sort, arrange and visualize the data in their web browser without having to write a line of code. Here the primary aim of the \reports{} is to lower the initial effort needed to explore the content of a datahub. However, expert users can further develop the code behind these reports opening the door to an even richer data exploration and visualization.

\subsection{Genome sequencing}
As a natural consequence of that genome sequencing gets cheaper by the time the amount of data produced is increasing as well. As a preprocessing step the raw data is typically aligned to a reference genome database in order to determine the most probable kinds of organisms present in the sequenced material \cite{mitocikk}. 
This digested data can be further analyzed e.g. for mutations.
With the evolution of genome sequencing equipment the data collected and processing tools also change and researchers have to come up with newer and newer evaluation workflows. \kooplex{} is and ideal platform to deploy next to a genome sequencing datahub, to explore and to develop such workflows, which may include the visualization of the results, as \reports. 
Alternatively, the visualization capabilities of the \kooplex{} platform can be extended with extra services, like the Cbioportal\footnote{\url{http://www.cbioportal.org/}}, which consumes data digested by \notebooks. 

\subsection{Kooplex for education}
\begin{figure}[h!]
    \centering
    \includegraphics[width=400px]{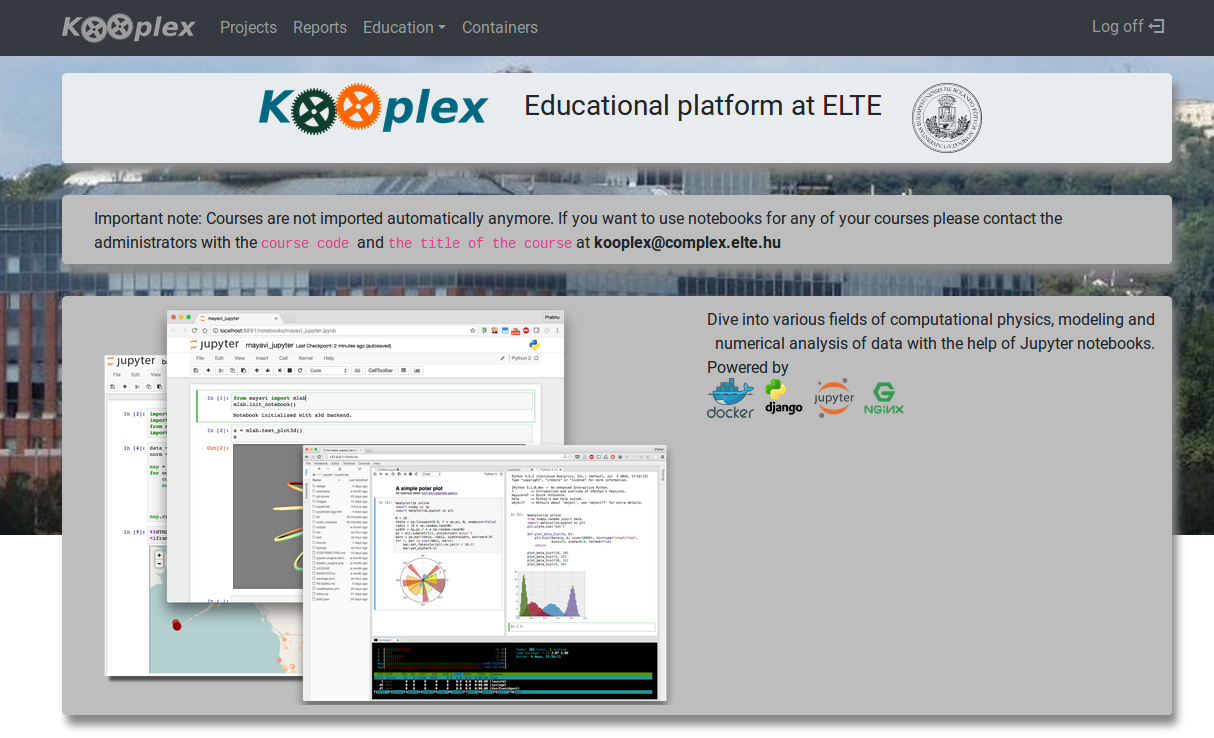}
    \caption{The front page of an instance of the \kooplex{} platform, that is used for  educational purposes at the Eötvös University at Budapest.}
    \label{fig:kooplex_edu}
\end{figure}

The teacher--student relationship can be regarded as a type of collaboration. 
With slight addition to the hub code a \kooplex-education instance has been operated for two years now at Eötvös University, Budapest\footnote{\url{https://kooplex-edu.elte.hu/hub}} (Fig. \ref{fig:kooplex_edu}.). 
A course taught at universities can easily mapped to the concept of \project, where by their role teachers are the owner or administrators of the course, whereas students act as collaborators with the restriction of having read-only access to the shared material. In order to extend functionalities of the \kooplex-education instance plugins are added to the hub to manage assignments, to schedule the collection of homework, to help correction and mechanisms to give feedback to students. Note, submitted assignments are analog to the \reports{} created from the \notebooks.

In \kooplex-education instance the directory structure is similar to Fig.~\ref{fig:filesystemlayout} with carefully applied access rights and it resembles after the one used in \emph{nbgrader}\footnote{\url{https://github.com/jupyter/nbgrader}}.
In this instance, depending on their role, whether a teacher or a student, personal working directories are the entry point for creating assignments or for working on them. 
The resource material, the common data or other auxiliary files needed for the completion of an assignments are placed read-only files in \emph{share} to avoid unnecessarily duplication.

\section{Summary}
\kooplex{} is a platform for easy access to datahubs, for collaborative work, for developing new workflows and for creating and publishing static or interactive reports.  
It is clear from the user feedback regarding the platform instances mentioned in \textbf{Section \ref{sec:kooplex-usecase}.} that the combination of such integrated services is attractive. Accessing the various modules in the same user space speeds up analysis, code development work and sharing, not having to move data and files around between disconnected components. 
The platform has been designed in a way that the integration of new tools taken up by the research community is straight forward. This feature helps to keep up with the ever evolving user requirements.

\section*{Acknowledgements}
This study has received funding from the European Union’s Horizon 2020 research and innovation program under Grant Agreement No. 643476 (COMPARE) and from National Research, Development and Innovation Fund of Hungary Project (FIEK\_16-1-2016-0005 to I.C.). The authors are grateful to G. Vattay, S. Laki and students at Eötvös University for help with thorough testing of the system.
This work was completed in the ELTE Excellence Program (783-3/2018/FEKUTSRAT) supported by the Hungarian Ministry of Human Capacities.
The work was supported by the Hungarian National
Research, Development and Innovation Office (NKFIH)
through Grants No. K120660, K109577, K124351,
K124152, KH129601, and the Hungarian Quantum
Technology National Excellence Program (Project No.
2017-1.2.1-NKP-2017- 00001).

\section*{Conflict of interest}
The authors declare that there is no conflict of interest regarding the publication of this manuscript.









\bibliographystyle{elsarticle-num-names}
\bibliography{ref}

\end{document}